\documentclass[final,5p,times,twocolumn]{elsarticle}

\usepackage{lipsum}
\usepackage{mathtools}
\usepackage{cuted}\usepackage{graphicx}
\usepackage{amssymb}
\usepackage{amsmath}
\usepackage[svgnames]{xcolor}
\usepackage{mathtools,slashed}
\usepackage[retainorgcmds]{IEEEtrantools}
\usepackage{physics}
\usepackage[compat=1.1.0]{tikz-feynhand}
\usepackage{tikz}
\usepackage{epstopdf}
\usepackage[utf8]{inputenc}
\usepackage{url}
\usepackage[colorlinks,citecolor=DarkGreen,linkcolor=DarkRed,urlcolor=DarkBlue]{hyperref}
\usepackage{newtxtext}
\usepackage{newtxmath}
\setcitestyle{square}

\journal{Physics Letters B}

\begin{document}

\begin{frontmatter}

\title{On the origin of the peak of the sound velocity for isospin imbalanced\\ strongly interacting matter}

\author[first]{Alejandro Ayala}
\author[second]{Bruno S. Lopes}
\author[second]{Ricardo L. S. Farias}
\author[first]{Luis C. Parra}

\affiliation[first]{organization={Instituto de Ciencias Nucleares, Universidad Nacional Autónoma de México},
            addressline={ Circuito Ext. S/N Ciudad Universitaria}, 
            city={CdMx},
            postcode={04510}, 
            state={CdMx},
            country={Mexico}}

\affiliation[second]{organization={Departamento de Física, Universidade Federal de
  Santa Maria},
            city={Santa Maria},
            postcode={RS 97105-900}, 
            state={Rio Grande do Sul},
            country={Brazil}}

\begin{abstract}
We study the properties of a system composed of strongly interacting matter with an isospin imbalance, using as an effective description of QCD the two-flavor Linear Sigma Model with quarks. From the one-loop effective
potential, including the two light quarks, pions and
sigma contributions, and enforcing the restrictions imposed by chiral symmetry, we show that the development of an isospin condensate comes together with the emergence of a Goldstone mode that provides a constraint for the chiral and isospin condensates as a result of a non-trivial mixing between the charged pions and the sigma. We compute the thermodynamical quantities of interest and in particular the sound velocity squared, showing that it presents a maximum for an isospin chemical potential similar to the one reported by lattice QCD results and also with a similar height. Therefore, we attribute the origin of the peak of the sound velocity to the proper treatment of the Goldstone mode and to the non-trivial mixing of the charged pions and sigma in the isospin condensed phase. 
\end{abstract}

\end{frontmatter}

\section{Introduction}
\label{introduction}

It is by now clear that Quantum Chromodynamics (QCD) possesses a rich phase structure, which has been brought to light from theoretical as well as experimental studies where strongly interacting matter is subject to the effects of finite temperature ($T$) and finite (baryon, strangeness, isospin) densities. Several avenues converge to shape up our current understanding on the subject; on the one hand, Lattice QCD (LQCD) calculations have revealed that the phase transition from a hadron to a quark-gluon plasma at zero baryon ($\mu_B$), strangeness ($\mu_s$) and isospin ($\mu_I$) chemical potentials is a smooth crossover that happens at a pseudocritical temperature of about 158 MeV~\cite{Borsanyi:2020fev}. Relativistic heavy-ion collisions have explored the phase diagram also in the $\mu_B$ -- $T$ plane, searching in particular for the possible existence of the critical end point, thought to exist as the end of a first order phase transition line, thus far with no conclusive results~\cite{Chen:2024zwk}. LQCD simulations cannot reliably explore deeper into the $\mu_B$ -- $T$ phase diagram due to the sign problem~\cite{Splittorff:2007ck,Hsu:2010zza,Aarts:2012yal,Nagata:2021ugx}. 

On the other hand, LQCD calculations with $\mu_I \neq 0,~\mu_B=\mu_s=0$ can be safely performed since they are not hindered by the sign problem. These calculations, together with theoretical considerations, have revealed the existence of a quark-antiquark superfluid phase for high enough $\mu_I$, that at $T=0$ is described by a second order transition from the hadron to the pion condensed phase at a critical $\mu_I^c$ given by the vacuum pion mass, $m_\pi$~\cite{Kogut:2002zg,Kogut:2002tm,Brandt:2016zdy,Brandt:2017zck,Brandt:2017oyy,Brandt:2018wkp,Son:2000xc,Son:2000by,Splittorff:2000mm,Cohen:2015soa,Lepori:2019vec,He:2005sp,Andersen:2007qv,Xia:2013caa,Abuki:2008wm,Khunjua:2018jmn,Khunjua:2018sro,Khunjua:2017khh,Ebert:2016hkd,Ueda:2013sia,Mannarelli:2019hgn}. Such calculations can be used as benchmarks to achieve a better understanding of the physical features of the computed thermodynamical variables, and in turn, effective models can be of utmost relevance for these purposes. In fact, recently, effective models~\cite{Andersen:2023ivj,Ayala:2023cnt,Lopes:2021tro,Adhikari:2019zaj,Avancini:2019ego,Andersen:2015eoa,Mukherjee:2006hq,Chen:2024cxh} have been used in the $T=0$, $\mu_B=0$ domain and have found a very good agreements with LQCD results~\cite{Brandt:2022fij,Brandt:2022hwy} for the behavior of thermodynamical quantities such as pressure, energy density and isospin density as functions of $\mu_I$. Finite temperature effects have also been explored~\cite{Stiele:2013pma,Adhikari:2018cea}. However, most of the models and/or perturbative calculations fail to reproduce the peak in the square of the sound velocity $c_s^2$, or require either a description in terms of a medium-dependent coupling~\cite{Ayala:2023mms} or the use of non-local interactions~\cite{Carlomagno:2024xmi,Carlomagno:2021gcy}. Depending on the implementation, LQCD calculations show the peak on $c_s^2$ within the range $1.5\ m_\pi \lesssim \mu_I \lesssim 2.6\ m_\pi$~\cite{Brandt:2022fij,Brandt:2022hwy,Abbott:2023coj,Abbott:2024vhj}. These calculations also find that the peak significantly exceeds the conformal limit $c_s^2 \leq 1/3$ for a wide range
of $\mu_I$ values. A larger than the conformal limit value for $c_s^2$ has also been inferred from recent analyses of neutron star data~\cite{Brandes:2023hma}

A crucial feature of a second order phase transition, signaled by the development of a condensate, is the spontaneous breaking of a continuous symmetry. In the present context, the broken symmetry is a $U(1)_I$ which must come together with the appearance of a Goldstone boson. Recall that when a system containing bosons enters a phase where Goldstone modes develop, a non-trivial rearrangement of the modes is required. In general, bosons mix and the normal modes do not correspond to the usual vacuum modes~\cite{He:2005nk,Carignano:2016lxe}. However, the condition for the system to possess a Goldstone mode can be identified from the mixing, provided the full boson content of the theory is accounted for. In strongly interacting systems, chiral symmetry dictates the boson content.

In this letter we show that the peak in the sound velocity for isospin imbalanced matter can be understood as resulting from the development of a Goldstone mode at the onset of the charged pion condensed phase. To obtain the result, we consider a simple scenario whereby chiral symmetry and its breaking are implemented in the two-flavor linear sigma model with quarks (LSMq), while we also allow the development of an isospin condensate. The model has the advantage of being renormalizable. This model has also been used in its two-- and three-- flavor versions in Refs.~\cite{Chiba:2023ftg,Kojo:2024sca}, respectively, where the peak of the sound velocity is found, although attributed to the effects induced by quarks for large values of $\mu_I$. Such result was obtained using a somewhat large boson self-coupling. Here instead, we use the relation between the couplings imposed by a Ward-Takahashi identity and show that the peak position and strength appears as a consequence of the proper description of the Goldstone mode in the condensed phase.

\section{Lineal Sigma Model with Quarks}
The Lagrangian for the LSMq is given by
\begin{eqnarray}
    \!\!\!\!\mathcal{L}&\!\!\!\!=\!\!\!\!& \frac{1}{2}\left[(\partial_{\mu}\sigma)^2+(\partial_{\mu}\vec{\pi})^2\right]+\frac{a^2}{2}(\sigma^2+\vec{\pi}^2)-\frac{\lambda}{4}(\sigma^2+\vec{\pi}^2)^2\nonumber \\
    &\!\!\!\!+\!\!\!\!&i \bar{\psi}\gamma^\mu \partial_{\mu} \psi- ig\bar{\psi} \gamma^{5} \vec{\tau} \cdot \vec{\pi} \psi -g \bar{\psi}\psi \sigma, 
\label{LSMqLagrangian}
\end{eqnarray}
where  
$\vec{\tau}=(\tau_{1},\tau_{2},\tau_{3})$
are the Pauli matrices, 
$\psi$ is a  $SU(2)_{L,R}$ fermion doublet, $\sigma$ is a real scalar field and $\vec{\pi}=(\pi_{1},\pi_{2},\pi_{3})$ is a triplet of real scalar fields, with $\pi_3$ corresponding to the neutral pion, whereas the charged ones are represented by the combinations
$\pi_{-}=(\pi_{1}+i\pi_{2})/\sqrt{2}$ and $\pi_{+}=(\pi_{1}-i\pi_{2})/\sqrt{2}$.
The parameters $a^2$, $\lambda$ and $g$ are real and positive definite. A conserved isospin charge can be added to the LSMq Hamiltonian, multiplied by the isospin chemical potential. The Lagrangian gets modified with the ordinary derivative becoming a covariant derivative
\begin{equation}
    \partial_{\mu} \to D_{\mu}= \partial_{\mu}+i\mu_{I} \delta_{\mu}^{0}, \quad  \partial^{\mu} \to D^{\mu}= \partial^{\mu}-i\mu_{I} \delta_{0}^{\mu},
\end{equation}
Because of the spontaneous breaking of the chiral symmetry, the $\sigma$ field acquires a non-vanishing vacuum expectation value
$\sigma\rightarrow \sigma + v.
$
An explicit chiral symmetry breaking term $-hv$ can be added to the Lagrangian, allowing the pions acquire a finite mass. For the present purposes, where we emphasize the dynamics of the pion fields, we work with the ansatz $\langle \bar\psi i\gamma_5\tau_3\psi\rangle=0$, together with $\langle \bar{u}i\gamma_5 d\rangle = \langle \bar{d}i\gamma_5 u\rangle^* \neq 0$~\cite{Mu:2010zz}. Then, the charged pion fields can be expanded from their condensates as $\pi_\pm\rightarrow \pi_\pm + \Delta\exp{\pm i\theta}/\sqrt{2}$, where the phase factor $\theta$ indicates the direction of the $U(1)_{I}$ symmetry breaking. The shift in the sigma field allows the quarks to develop a mass given by $m_f=g v$.

In the condensed phase the tree-level potential is
\begin{equation}
    V_{tree}=-\frac{a^2}{2}\left(v^{2}+\Delta^2\right)+\frac{\lambda}{4}\left(v^2+\Delta^2 \right)^2-\frac{1}{2}\mu_I^2\Delta^2-hv.
\label{treeeffectivepotential}
\end{equation}
\subsection{One-loop effective potential}
After the breaking of the chiral and isospin symmetries, the one-loop effective potential contains, in addition to the interaction terms, the quadratic terms, both in the fermion and the boson sectors. These terms are no longer diagonal in flavor space. As a consequence, the $u$ and $d$ quarks and the pions and sigma fields mix, respectively, producing non-trivial dispersion relations in each case.
The fermion contribution to the one-loop effective potential is given by
\begin{equation}
V^1_{f} = i\int\frac{d^4k}{(2\pi)^4} \ln\left(\det\left\{S_f^{-1}\right\}\right)
\label{fermeffpot}
\end{equation}
where $S_f^{-1}$ is the inverse quark propagator, a non-diagonal $2\times 2$ matrix in flavor space. It can be shown~\cite{Ayala:2023cnt} that 
\begin{equation}
V_{f}^1= -2N_c\int\frac{d^3k}{(2\pi)^3}\left[E_\Delta^u +E_\Delta^d\right],
\label{FermionOneLoopCorrection}
\end{equation}
where $N_c=3$ is the number of colors and
\begin{eqnarray}
E_\Delta^{\stackrel{d}{u}} = \left\{\left(\sqrt{k^2+m_f^2}\mp\mu_I/2\right)^2+g^2\Delta^2\right\}^{1/2},
\label{fermionenergies}
\end{eqnarray}
where we chose that
$\mu_d=\mu_I/2$ and $\mu_u=-\mu_I/2$.
The boson contribution to the one-loop effective potential is given by 
\begin{equation}
V^1_{b} = -\frac{i}{2}\int\frac{d^4k}{(2\pi)^4} \ln\left(\det\left\{D_b^{-1}\right\}\right)
\label{BosonOneLoopCorrection}
\end{equation}
where $D_b^{-1}$ is the non diagonal boson inverse propagator which can be written as
\begin{strip}
\begin{equation}
D_b^{-1}=\begin{pmatrix}
        K^2 -m^2_{\sigma} & -\sqrt{2} \lambda v \Delta e^{-i \theta} & -\sqrt{2} \lambda v \Delta e^{i \theta} & 0 \\
        -\sqrt{2} \lambda v \Delta e^{i \theta} & K^2 -m^2_{ch} +\mu_I^2 + 2 \mu_I k_0 & -\lambda \Delta^2 e^{2i \theta} &  0 \\
        -\sqrt{2} \lambda v \Delta e^{-i \theta}& -\lambda \Delta^2 e^{-2i \theta} & K^2 -m^2_{ch} +\mu_I^2 - 2 \mu_I k_0 &  0 \\
        0 & 0 & 0 & K^2 -m^2_{\pi_0}\\
\end{pmatrix}
\label{InversePropagator}
\end{equation}
\end{strip}
with $K^2=k_0^2-k^2$ and where we have used the shorthand notation 
\begin{eqnarray}
    m_\sigma^2 &=& \lambda(3 v^2+\Delta^2)-a^2\nonumber\\
    m_{\pi_0}^2 &=& \lambda(v^2+\Delta^2)-a^2\nonumber\\
    m_{ch}^2 &=& \lambda(v^2+2\Delta^2)-a^2.
\label{DinamicalMass}
\end{eqnarray}
Notice that the structure of the boson inverse propagator, Eq.~(\ref{InversePropagator}), makes it evident that the only boson that does not mix with the rest is the neutral pion. The charged pions and the sigma do mix. In this sense, only $m_{\pi_0}^2$, defined in Eqs.~(\ref{DinamicalMass}), corresponds to the square of the neutral pion mass, whereas $m_\sigma^2$ and $m_{ch}^2$ are not the squares of the sigma and charged pion masses, but rather only useful combinations of the parameters that appear in the analysis. However, in the absence of mixing, they would represent the square of the corresponding particle masses.

\subsection{Goldstone mode condition}

The breaking of the $U(1)_{I}$ global symmetry comes together with the development of a Goldstone boson. To find the restrictions imposed on $v$ and $\Delta$ by the development of a massless mode, we need to look at the limit where the components of $K^\mu\to 0$ and set the determinant of the inverse boson propagator to zero. This is equivalent to keeping only the product of the masses in the calculation of this determinant. Following this procedure, we can identify the masses of the excitations in the isospin broken phase as corresponding to the explicit factors that make up the determinant. We thus need to find the solutions for $\Delta$ satisfying 
\begin{eqnarray}
    \!\!\!\!\!\!\!\!\!\!\!\mathop{\det{D_b^{-1}}}_{K^\mu\to 0}&\!\!\!\!=\!\!\!\!& m_{\pi^0}^2 m_{\sigma}^2 \!\!\left(m_{\pi^0}^2 \!\! - \!\mu_I^2\right)\!\!\left(m_{\pi^0}^2 \! + 2 \Delta ^2 \lambda \frac{ m_{\pi^0}^2}{m_{\sigma}^2} \! - \! \mu_I^2\right)\!\!=\! 0,
\label{ProductOfMasses}
\end{eqnarray}
which, since $m_{\pi^0}^2$ and $m_\sigma^2$ are positive definite, is equivalent to finding the solutions for
\begin{equation}
    \left(m_{\pi^0} - \mu_I\right) \left(\sqrt{m_{\pi^0}^2 + 2 \Delta ^2 \lambda \frac{ m_{\pi^0}^2}{m_{\sigma}^2}} - \mu_I\right) = 0
    \label{GoldstoneModeCondition}
\end{equation}
There are three possible solutions out of which only two are real on the whole $\mu_I\geq m_\pi$ domain, these are
\begin{eqnarray}
    \!\!\!\!\!\!\!\!\!\!\Delta_1 &\!\!\!\!=\!\!\!\!& \sqrt{\frac{\mu_I^2 - 2 \left(3 \lambda v^2 - 2 a^2\right) + \sqrt{4 a^4 + 4 \mu_I^2 \left(6 \lambda v^2 - a^2\right) + \mu_I^4}}{6 \lambda}}\nonumber\\ 
    \!\!\!\!\!\!\!\!\!\!\Delta_2 &\!\!\!\!=\!\!\!\!& \sqrt{\frac{\mu_I^2 - (\lambda v^2 - a^2)}{\lambda}}.
\label{Mode3DeltaCondentate}
\end{eqnarray}

Remarkably, $\Delta_2$ corresponds to the equation obtained from the $\Delta$-gap equation at tree-level. Since at this level, this equation coincides with the Goldstone mode condition, then the peak in the sound velocity should be already present when using the tree-level solutions for the gap equations. As we proceed to show, this is indeed the case. Before embarking on this discussion, we first go back to the full one-loop result and explore the effect of considering the $\Delta_1$ solution, which comes from the full-fledged treatment of the particle content of the model arising from the non-trivial mixing induced by the one-loop correction.

Equations~(\ref{fermeffpot}) and~(\ref{BosonOneLoopCorrection}) contain ultraviolet divergences that can be isolated to write
\begin{eqnarray}
V^1_f&=&V^{1 (vacuum)}_f + V^{1 (matter)}_f\nonumber\\
V^1_b&=&V^{1 (vacuum)}_b + V^{1 (matter)}_b
\end{eqnarray}
where the first terms on the right-hand side contain the divergences and the second ones are finite and are correspondingly dubbed {\it vacuum} and {\it matter} pieces, respectively. 
\begin{figure}[t]
    \centering
    \includegraphics[width=\linewidth]{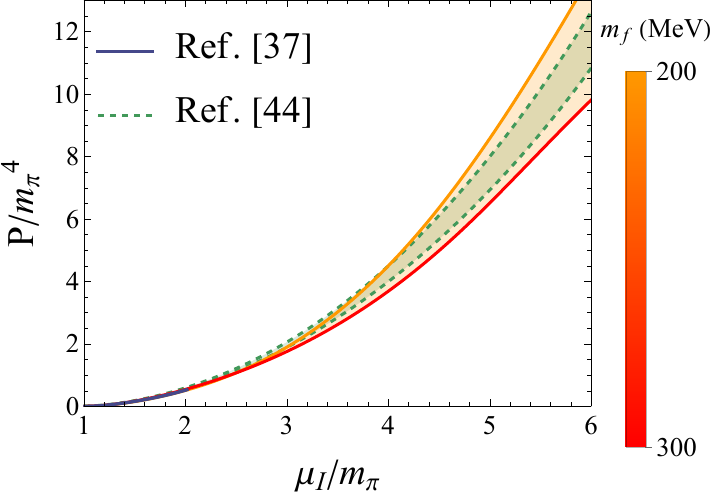}
    \caption{Pressure as a function of $\mu_I/m_\pi$ for a range of the quark mass 200 MeV $< m_f < 300$ MeV,  which requires 424 MeV $< m_\sigma < 616$ MeV. For comparison, the LQCD data from a private communication with the authors of Ref.~\cite{Brandt:2022hwy} (gray) and~\cite{Abbott:2024vhj} (green), are also shown.}
    \label{fig:PressPlot}
\end{figure}
\subsection{Renormalization}

We work in the $\overline{\mbox{MS}}$ scheme and carry out the renormalization procedure in two steps. First, we set $\Delta=0$ to find the counter-terms in the chirally broken symmetry, isospin symmetric phase. This is accomplished by adding the tree and vacuum pieces of the effective potential and imposing the stability conditions which require that, along the $v$-direction, the position of the minimum of the tree-level potential is not changed with the addition of the vacuum pieces, neither the curvature at the minimum, which is given by the vacuum sigma mass. The Ward-Takahashi identity\begin{equation}
    D^{-1}_{\sigma}-D^{-1}_{\pi}=-2\lambda v^2,
    \label{WardIdentity}
\end{equation}
which reflects the partial conservation of axial current, can be used to ensure that both counter-terms are in fact equal. Next, we allow $\Delta\neq 0$. In this phase, there is an extra divergence proportional to $\Delta^2$ that needs to be cured by the addition of a counter-term. To fix this counter-term, we require that the second derivative in the $\Delta$-direction vanishes. This condition is equivalent to requiring the vanishing of the mass of a boson excitation, and thus that for $\Delta\neq 0$ the system contains the Goldstone mode. The calculation of the matter pieces needs to be carried out numerically. In order to ensure that these contributions produce a continuous $\Delta$ condensate at the interface $\mu_I=m_\pi$, a finite constant part needs to be added to the counter-term that renormalizes the $\Delta^2$ divergence. 

For each value of $\mu_I\geq m_\pi$, the solution $\Delta_1$ needs to be enforced for the system to be in the Goldstone mode. The second relation that is needed to determine both condensates as functions of $\mu_I$ is obtained finding the value of $v$ that minimizes the renormalized one-loop effective potential. With these solutions for the condensates at hand, we compute the  normalized pressure as
\begin{eqnarray}
    \!\!\!\!\!\!\!\!P(v,\Delta_1(v),\mu_I) \!=\! - \left[V(v,\Delta_1(v),\mu_I)-V(f_\pi,0,m_{\pi})\right]\!.
\end{eqnarray}
To fix the parameters, notice that chiral symmetry, encoded in the Ward-Takahashi identity of Eq.~(\ref{WardIdentity}), can also be used to show that the couplings and the vacuum masses in the model are related by
\begin{eqnarray}
    2 g^2=\lambda,
    \quad    
    m_\sigma^2=4\, m_f^2 + m_{\pi}^2.
    \label{MassesRelation}
\end{eqnarray}
\begin{figure}[t]
    \centering
    \includegraphics[width=0.85\linewidth]{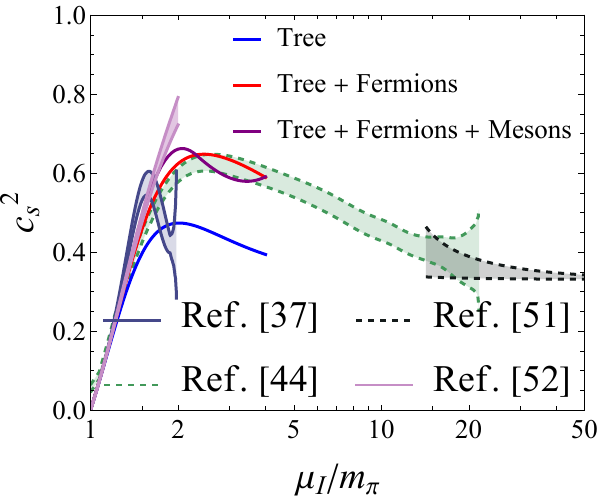}
    \caption{Square of the sound velocity obtained from computing the condensates $v$ and $\Delta$ using only the tree-level potential (blue line), the tree-level plus the one-loop fermion contribution (red line) and finally the tree-level, the one-loop fermion and the one-loop boson contributions (purple). For the calculations, the quark mass is taken as $m_f=240$ MeV. For comparison, the LQCD data from a private communication with the authors of Ref.~\cite{Brandt:2022hwy} (gray) and~\cite{Abbott:2024vhj} (green), the pQCD result from  Ref.~\cite{Fujimoto:2023mvc} (black) and the $\chi$PT result from Ref.~\cite{Adhikari:2019mdk} (light purple) are also shown.}
    \label{fig:sound-velocity-per-species}
\end{figure}

Since the vacuum quark mass is obtained as $m_f=gf_\pi$, then for fixed vacuum pion mass and decay constant the model has one free parameter, which can be chosen as the vacuum quark mass. We take $m_\pi=140$ MeV and $f_\pi=92.4$ MeV. The thermodynamical functions of interest can be obtained from the pressure as
\begin{eqnarray}
    n_I &=& \partial P/\partial \mu_I,\quad
    \varepsilon=-P +\mu_I n_I,\quad
    c_s^2=\partial P/\partial \varepsilon.
\end{eqnarray}

\begin{figure}[t]
    \centering
    \includegraphics[width=\linewidth]{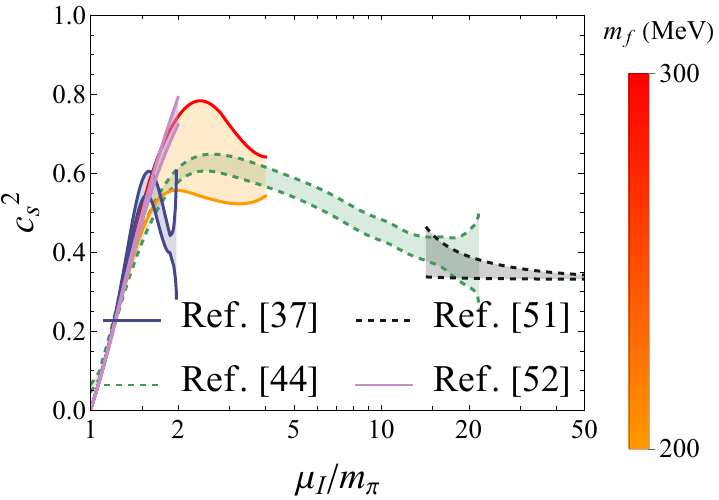}
    \caption{Square of the sound velocity for a range of the quark mass 200 MeV $< m_f < 300$ MeV, which requires 424 MeV $< m_\sigma < 616$ MeV. For comparison, the LQCD calculations from Ref.~\cite{Brandt:2022hwy} (gray) and Ref.~\cite{Abbott:2024vhj} (green), the pQCD result from  Ref.~\cite{Fujimoto:2023mvc} (black) and the $\chi$PT result from Ref.~\cite{Adhikari:2019mdk} (light purple) are also shown. Notice that the peak position (1.9 $m_\pi<\mu_I<2.6\, m_\pi$) is somewhat in between the two LQCD data sets and gets displaced  toward larger values of $\mu_I$ and increases its height as $m_f$ increases.}
    \label{fig:cs2Plot}
\end{figure}

Figure~\ref{fig:PressPlot} shows the pressure as a function of $\mu_I/m_\pi$. The calculation is performed using a range of values $200$ MeV $< m_f < 300$ MeV which requires $424$ MeV $< m_\sigma < 616$ MeV. For comparison, the LQCD calculation from Refs.~\cite{Brandt:2022hwy,Abbott:2024vhj} is also shown. To show that the tree-level solution, $\Delta_2$, already contains the seed for the peak of the sound velocity, Fig.~\ref{fig:sound-velocity-per-species} shows the square of this quantity as a function of $\mu_I/m_\pi$, for a single value of $m_f=240$ MeV. The figure also shows the square of the sound velocity obtained when considering successively the one-loop contribution, first from only fermions, and then from the full particle content of the model. Notice that the peak evolves as the different contributions kick in. A peak of a modest strength and the correct shape is present at tree-level. The strength increases and the shape remains when considering the tree-level plus fermion contributions. When including the full-fledged one-loop content of the theory, the peak sharpens, preserving its intensity and becoming located somewhere in between the two LQCD results, also shown in the figure.  Figure~\ref{fig:cs2Plot} shows the square of the sound velocity, also as a function of $\mu_I/m_\pi$, for the same range of masses $m_f$ and $m_\sigma$ as in Fig.~\ref{fig:PressPlot}, together with the corresponding LQCD results from Refs.~\cite{Brandt:2022hwy,Abbott:2024vhj}. Notice that the square of the sound velocity reaches a maximum, comparable to the LQCD maximum, which gets displaced toward larger values of $\mu_I$ and increases its height as $m_f$ increases. For intermediate values of $m_f$ in the considered range, the location of the peak is again somewhat in between the two LQCD sets. As a guide from known limits of the square of the sound velocity, in the low and high $\mu_I$ limits, Figs.~\ref{fig:sound-velocity-per-species} and~\ref{fig:cs2Plot} also show results from chiral perturbation theory from Ref.~\cite{Adhikari:2019mdk} and from a perturbative QCD calculation from Ref.~\cite{Fujimoto:2023mvc}, including the effects of a pairing gap, respectively.

\section{Summary and Conclusions}

In summary, we have studied the properties of a system composed of strongly interacting matter with an isospin imbalance. To this end, we have resorted to an effective QCD description using a two-flavor LSMq, computing the effective potential up to one-loop order, including quarks, pions and sigma and enforcing the restrictions imposed by chiral symmetry, which determine, up to the choice of the mass of the quarks, the parameters of the model. We have used the renormalizability of the model to find the one-loop counterterms under suitable conditions, called the stability conditions, namely, the requirement that the tree-level plus one-loop effective potential minimum and curvature at the minimum, in the $v$ direction, for $\Delta=0$, remain at the values $f_\pi$ and $m_\sigma^2$, respectively. Renormalization ensures that the vacuum corrections are totally under control.

We emphasize that the calculation employs the complete set of fermions and bosons required by chiral symmetry in the $SU(2)$ version of the model. In the boson sector, this corresponds to accounting for the three pions and the sigma field. The sigma field, which is usually not included in this kind of analyses, plays a crucial role for the description of the condensed phase due to its non-trivial mixing with the charged pions. The analysis shows that one of the conditions for the presence of a Goldstone mode corresponds to the $\Delta$-condensate gap equation at tree-level. Therefore, the seed of the peak in the sound velocity should be already present at tree-level, as is the case. Adding successively the one-loop contributions from fermions and bosons increases the strength and sharpens the peak, respectively, making its location to be somewhere in between the two LQCD results used for comparison. 

The two-flavor LSMq has of course limited accuracy and, as such, we can only expect to obtain out of it a ballpark description. Nevertheless, it is also simple enough such that interesting features of the system where the model applies can be revealed. One of these is the way the charged pion condensed phase develops. We have shown that this comes together with the development of a Goldstone mode and a nontrivial mixing between the charged pions and the sigma. From the one-loop effective potential, the pressure and the rest of the thermodynamical variables of interest can be computed. Our results show a range of validity for $m_\pi \leq \mu_I \lesssim 4 \, m_\pi$, signaled by the development of a maximum of the energy density, which indicates the need to include extra degrees of freedom in the description. This is to be expected since, as pointed out in Ref.~\cite{Son:2000xc}, when more pions are forced into the condensed phase, they become closely packed, and their interaction becomes stronger, opening the possibility of forming other states such as $\rho$'s, which can then contribute to the pressure and energy density, which sets a limit of applicability of the model to the domain $\mu_I\lesssim m_\rho$, since it does not contain interactions with $\rho$'s. This point is for the moment of a speculative nature and is being discussed in a longer version of the present work, were we also make a systematic study of the parameter space, provide more details of the calculation and present the analysis of the properties of the thermodynamical variables of interest. This work is currently in preparation and will be reported elsewhere.

\section*{Acknowledgements}
Support for this work was received in part by: UNAM-PAPIIT grant number IG100322; Consejo Nacional de Humanidades, Ciencia y Tecnolog\'{\i}a grant number CF-2023-G-433; Conselho Nacional de Desenvolvimento Científico e Tecnol\'ogico (CNPq) grant numbers 312032/2023-4 (R.L.S.F.) and 141270-2023-3 (B.S.L.); Fundaç\~ao de Amparo \'a
Pesquisa do Estado do Rio Grande do Sul (FAPERGS): grant numbers
19/2551-0000690-0 and 19/2551-0001948-3 (R.L.S.F.); The work is part of the project Instituto Nacional de Ci\^encia e Tecnologia — F\'{\i}sica Nuclear e
Aplicaç\~oes (INCT–FNA), grant number 464898/2014-5.

\appendix

\section{Vacuum-Matter Separation}
Both the fermion and boson one-loop contributions to the effective potential exhibit a highly intricate structure. In particular, their ultraviolet divergences preclude straightforward numerical integration and therefore need to be isolated first. We accomplish this by adding and subtracting a term to the integrands such that
\begin{equation}
    \sum_i E_i + f(k,v,\Delta,\mu_I) - f(k,v,\Delta,\mu_I),
\end{equation}
where $E_i$ represent the eigenvalues of the corresponding (fermion or boson) determinant. To capture the ultraviolet structure, the function $f$ is chosen from an asymptotic interpolation of the form
\begin{equation}
    f = \sum_{j=1}^{\infty} a_j (k^2 + m_j^2)^{(3-2j)/2},
    \label{expansion}
\end{equation}
and we compute the divergent parts using dimensional regularization. Notice that the polynomial expansion cannot be truncated below the order $(3-2j)/2 \geq -d$. For the case of interest, with $d=3$, selecting $j \leq (3+2d)/2 = 9/2$ ensures the convergence of the integrals. However, more terms could in principle be added to the polynomial, Eq.~(\ref{expansion}), to further improve the convergence of the remaining integral. By these means, we identify two terms contributing to the integrals of the energy eigenvalues. The first, corresponding to the integral of the asymptotic expansion $f(k,v,\Delta,\mu_I)$, which we call the {\it vacuum} piece, contains the original divergences and can be integrated using dimensional regularization. The integrand of the second term, that we call the {\it matter} piece, is proportional to $\sum_i E_i - f(k,v,\Delta,\mu_I)$ and is finite, making it numerically tractable.

Applying this vacuum extraction technique to fermions, we use
\begin{equation}
    f_f = 2 \sqrt{k^2 + M_f^2} + \frac{\Delta^2 g^2 \mu_I^2}{(k^2 + M_f^2)^{3/2}} + \frac{\Delta^2 g^2 \mu_I^4}{(k^2 + M_f^2)^{5/2}},
\end{equation}
with $M_f^2 = g^2 (v^2 + \Delta^2)$. For bosons, we use
\begin{align}
    f_b = & \sqrt{k^2 + m_\sigma^2} + 2\sqrt{k^2 + m_{ch}^2} - \frac{\Delta^2 \lambda^2 (\Delta^2 + 4 v^2)}{4 (k^2 + m_{ch}^2)^{3/2}} \nonumber\\
    & + \frac{\Delta^2 \lambda^2 [(v^2 + \Delta^2) \mu_I^2 + 6 \lambda v^4]}{4 (k^2 + m_{ch}^2)^{5/2}}.
\end{align}

\section{Renormalization}
Once the divergences are analytically extracted, the renormalization procedure can be implemented. We define the field renormalization factors as follows:
\begin{eqnarray}
    \psi \rightarrow Z_{f}^{-1/2} \psi, \quad
    \sigma \rightarrow Z_{b}^{-1/2} \sigma, \quad
    \vec{\pi} \rightarrow Z_{b}^{-1/2} \vec{\pi}.
\end{eqnarray}
This implies that the model parameters must also be renormalized:
\begin{eqnarray}
    a^2 \rightarrow Z_{a} a^2, \quad
    \lambda \rightarrow Z_{\lambda} \lambda, \quad
    g \rightarrow Z_{g} g.
\end{eqnarray}
The Ward-Takahashi identity for axial current conservation is reflected in the relation $Z_{\lambda} = Z_{a}$. Defining $Z_a = 1 - \delta_a$, $Z_\lambda = 1 - \delta_\lambda$, and $Z_b = 1 - \delta$, the tree-level potential in Eq.~(\ref{treeeffectivepotential}) is renormalized by adding the counter-terms
\begin{equation}
    \delta V_{tree} = \frac{\delta_a a^2}{2} \left( v^{2} + \Delta^2 \right) - \frac{\delta_\lambda \lambda}{4} \left( v^2 + \Delta^2 \right)^2 + \frac{\delta}{2} \mu_I^2 \Delta^2.
\label{tadpolespotential}
\end{equation}

We determine the three counter-terms by enforcing the stability conditions and ensuring the continuity of the one-loop effective potential solutions at $\mu_I = m_\pi$. This is achieved by using the conditions:
\begin{equation}
    \mathop{\frac{\partial V_{eff}}{\partial v}}_{\mu_I \to m_\pi} = 0, \quad \mathop{\frac{\partial^2 V_{eff}}{\partial v^2}}_{\mu_I \to m_\pi} = m_\sigma^2, \quad \mathop{\frac{\partial^2 V_{eff}}{\partial \Delta^2}}_{\mu_I \to m_\pi} = 0.
    \label{stability}
\end{equation}

In general, each counter-term has a divergent and a finite piece. From the Ward-Takahashi identity, Eq.~(\ref{WardIdentity}), one readily obtains $\delta_a = \delta_{\lambda}$. It can be shown that the divergent counter-term parts are
\begin{eqnarray}
    \delta_{a,div} = \frac{3 \lambda}{8 \pi^2 \epsilon}, \quad \delta_{\lambda,div} = \frac{3 \lambda - 2 g^4 N_c / \lambda}{4 \pi^2 \epsilon},
\end{eqnarray}
and therefore, from the Ward-Takahashi identity one gets
\begin{equation}
    4 g^4 N_c = 3 \lambda^2,
\end{equation}
which represents a direct relation between the boson-fermion coupling and the boson self-coupling. Also, at tree level, we can express these couplings as functions of the model parameters
\begin{equation}
    \lambda = \frac{m_\sigma^2 - m_{\pi}^2}{2 f_\pi^2}, \quad g = \frac{m_f}{f_\pi} ,
\end{equation}
where the masses refer to their values in the vacuum. For $N_c = 3$, this leads to the non-trivial mass relation
\begin{equation}
    m_\sigma^2 = 4\, m_f^2 + m_{\pi}^2.
\end{equation}

The renormalization procedure ensures that the theory remains consistent and finite while respecting the symmetries of the model. This mass relation, derived from the Ward-Takahashi identity, plays a crucial role in connecting the low-energy effective theory with the underlying microscopic parameters, providing an essential check for the consistency of the model at both the tree and the one-loop levels.

\section{Matter Contribution}

The one-loop contribution also contains the finite matter part. It is important to enforce that the stability conditions, Eqs.~(\ref{stability}), are not altered by the inclusion of the matter contribution which needs to be numerically computed. In particular, we can show that the two derivatives with respect to \(v\) are indeed independently satisfied. However, the condition for the second derivative with respect to \(\Delta\) is not directly satisfied, and this happens both for fermions and bosons. This issue is resolved by adding a finite counter-term of the form \(\delta \Delta^2 \mu_I^2 / 2\). The definition of \(\delta\) is given by the expression:
\begin{align}
    \delta &= -\frac{1}{m_\pi^2}\mathop{\int^{\infty}_0 \frac{d^3k}{(2\pi)^3} \frac{\partial^2}{\partial \Delta^2} \left( V^1_{f} - f_f(k, v, \Delta, \mu_I) \right)}_{\mu_I \to m_\pi}\nonumber \\
    &\quad - \frac{1}{m_\pi^2}\mathop{\int^{\infty}_0 \frac{d^3k}{(2\pi)^3} \frac{\partial^2}{\partial \Delta^2} \left( V^1_{b} - f_b(k, v, \Delta, \mu_I) \right)}_{\mu_I \to m_\pi}.
\end{align}

After fixing this additional counter-term, we are ready to proceed with the numerical exploration for \(\mu_I > m_\pi\). To this end, we use a step size of \(m_\pi / 100\). The computation is performed independently for the two \(\Delta\) solutions, given in Eq.~(\ref{Mode3DeltaCondentate}). Since the potential energy is always smaller for $\Delta_1$ than for $\Delta_2$, we keep only the former and discard the latter. At each step in \(\mu_I\), we search for the values of \(v\) that minimize the one-loop effective potential and to compute the pressure, and from it, the square of the sound velocity.

The error in the solutions is estimated taking the local standard deviation of the data. For the pressure this is less than \(0.125\%\), however, the cumulative error in \(c_s^2\) increases as $\mu_I$ increases and it is limited to \( 1.638\%\) in the peak region and to \( 4.880\%\) in the farthest plotted region. For larger values of $\mu_I$ (not shown in the plots), the error increases to about $10\%$. It is therefore important to tame this error, refining the numerical calculation for further analyses, when extending the region of $\mu_I$ for the computation of the thermodynamical variables of interest, in particular to explore the consequences of the findings of this work in the asymptotic domain when including extra degrees of freedom such as interactions with $\rho$'s.

\bibliography{biblio}

\providecommand{\noopsort}[1]{}\providecommand{\singleletter}[1]{#1}%
\begin{thebibliography}{10}
\expandafter\ifx\csname url\endcsname\relax
  \def\url#1{\texttt{#1}}\fi
\expandafter\ifx\csname urlprefix\endcsname\relax\def\urlprefix{URL }\fi
\expandafter\ifx\csname href\endcsname\relax
  \def\href#1#2{#2} \def\path#1{#1}\fi

\bibitem{Borsanyi:2020fev}
S.~Borsanyi, Z.~Fodor, J.~N. Guenther, R.~Kara, S.~D. Katz, P.~Parotto, A.~Pasztor, C.~Ratti, K.~K. Szabo, {QCD Crossover at Finite Chemical Potential from Lattice Simulations}, Phys. Rev. Lett. 125~(5) (2020) 052001.
\newblock \href {http://arxiv.org/abs/2002.02821} {\path{arXiv:2002.02821}}, \href {https://doi.org/10.1103/PhysRevLett.125.052001} {\path{doi:10.1103/PhysRevLett.125.052001}}.

\bibitem{Chen:2024zwk}
J.~Chen, et~al., {Properties of the QCD Matter -- An Experimental Review of Selected Results from RHIC BES Program} (July 2024).
\newblock \href {http://arxiv.org/abs/2407.02935} {\path{arXiv:2407.02935}}.

\bibitem{Splittorff:2007ck}
K.~Splittorff, J.~J.~M. Verbaarschot, {The QCD Sign Problem for Small Chemical Potential}, Phys. Rev. D 75 (2007) 116003.
\newblock \href {http://arxiv.org/abs/hep-lat/0702011} {\path{arXiv:hep-lat/0702011}}, \href {https://doi.org/10.1103/PhysRevD.75.116003} {\path{doi:10.1103/PhysRevD.75.116003}}.

\bibitem{Hsu:2010zza}
S.~D.~H. Hsu, D.~Reeb, {On the sign problem in dense QCD}, Int. J. Mod. Phys. A 25 (2010) 53--67.
\newblock \href {https://doi.org/10.1142/S0217751X10047968} {\path{doi:10.1142/S0217751X10047968}}.

\bibitem{Aarts:2012yal}
G.~Aarts, {Complex Langevin dynamics and other approaches at finite chemical potential}, PoS LATTICE2012 (2012) 017.
\newblock \href {http://arxiv.org/abs/1302.3028} {\path{arXiv:1302.3028}}.

\bibitem{Nagata:2021ugx}
K.~Nagata, {Finite-density lattice QCD and sign problem: Current status and open problems}, Prog. Part. Nucl. Phys. 127 (2022) 103991.
\newblock \href {http://arxiv.org/abs/2108.12423} {\path{arXiv:2108.12423}}, \href {https://doi.org/10.1016/j.ppnp.2022.103991} {\path{doi:10.1016/j.ppnp.2022.103991}}.

\bibitem{Kogut:2002zg}
J.~B. Kogut, D.~K. Sinclair, {Lattice QCD at finite isospin density at zero and finite temperature}, Phys. Rev. D 66 (2002) 034505.
\newblock \href {http://arxiv.org/abs/hep-lat/0202028} {\path{arXiv:hep-lat/0202028}}, \href {https://doi.org/10.1103/PhysRevD.66.034505} {\path{doi:10.1103/PhysRevD.66.034505}}.

\bibitem{Kogut:2002tm}
J.~B. Kogut, D.~K. Sinclair, {Quenched lattice QCD at finite isospin density and related theories}, Phys. Rev. D 66 (2002) 014508.
\newblock \href {http://arxiv.org/abs/hep-lat/0201017} {\path{arXiv:hep-lat/0201017}}, \href {https://doi.org/10.1103/PhysRevD.66.014508} {\path{doi:10.1103/PhysRevD.66.014508}}.

\bibitem{Brandt:2016zdy}
B.~B. Brandt, G.~Endrodi, {QCD phase diagram with isospin chemical potential}, PoS LATTICE2016 (2016) 039.
\newblock \href {http://arxiv.org/abs/1611.06758} {\path{arXiv:1611.06758}}, \href {https://doi.org/10.22323/1.256.0039} {\path{doi:10.22323/1.256.0039}}.

\bibitem{Brandt:2017zck}
B.~B. Brandt, G.~Endrodi, S.~Schmalzbauer, {QCD at finite isospin chemical potential}, EPJ Web Conf. 175 (2018) 07020.
\newblock \href {http://arxiv.org/abs/1709.10487} {\path{arXiv:1709.10487}}, \href {https://doi.org/10.1051/epjconf/201817507020} {\path{doi:10.1051/epjconf/201817507020}}.

\bibitem{Brandt:2017oyy}
B.~B. Brandt, G.~Endrodi, S.~Schmalzbauer, {QCD phase diagram for nonzero isospin-asymmetry}, Phys. Rev. D 97~(5) (2018) 054514.
\newblock \href {http://arxiv.org/abs/1712.08190} {\path{arXiv:1712.08190}}, \href {https://doi.org/10.1103/PhysRevD.97.054514} {\path{doi:10.1103/PhysRevD.97.054514}}.

\bibitem{Brandt:2018wkp}
B.~B. Brandt, G.~Endrodi, S.~Schmalzbauer, {QCD at nonzero isospin asymmetry}, PoS Confinement2018 (2018) 260.
\newblock \href {http://arxiv.org/abs/1811.06004} {\path{arXiv:1811.06004}}, \href {https://doi.org/10.22323/1.336.0260} {\path{doi:10.22323/1.336.0260}}.

\bibitem{Son:2000xc}
D.~T. Son, M.~A. Stephanov, {QCD at finite isospin density}, Phys. Rev. Lett. 86 (2001) 592--595.
\newblock \href {http://arxiv.org/abs/hep-ph/0005225} {\path{arXiv:hep-ph/0005225}}, \href {https://doi.org/10.1103/PhysRevLett.86.592} {\path{doi:10.1103/PhysRevLett.86.592}}.

\bibitem{Son:2000by}
D.~T. Son, M.~A. Stephanov, {QCD at finite isospin density: From pion to quark - anti-quark condensation}, Phys. Atom. Nucl. 64 (2001) 834--842.
\newblock \href {http://arxiv.org/abs/hep-ph/0011365} {\path{arXiv:hep-ph/0011365}}, \href {https://doi.org/10.1134/1.1378872} {\path{doi:10.1134/1.1378872}}.

\bibitem{Splittorff:2000mm}
K.~Splittorff, D.~T. Son, M.~A. Stephanov, {QCD - like theories at finite baryon and isospin density}, Phys. Rev. D 64 (2001) 016003.
\newblock \href {http://arxiv.org/abs/hep-ph/0012274} {\path{arXiv:hep-ph/0012274}}, \href {https://doi.org/10.1103/PhysRevD.64.016003} {\path{doi:10.1103/PhysRevD.64.016003}}.

\bibitem{Cohen:2015soa}
T.~D. Cohen, S.~Sen, {Deconfinement Transition at High Isospin Chemical Potential and Low Temperature}, Nucl. Phys. A 942 (2015) 39--53.
\newblock \href {http://arxiv.org/abs/1503.00006} {\path{arXiv:1503.00006}}, \href {https://doi.org/10.1016/j.nuclphysa.2015.07.018} {\path{doi:10.1016/j.nuclphysa.2015.07.018}}.

\bibitem{Lepori:2019vec}
L.~Lepori, M.~Mannarelli, {Multicomponent meson superfluids in chiral perturbation theory}, Phys. Rev. D 99~(9) (2019) 096011.
\newblock \href {http://arxiv.org/abs/1901.07488} {\path{arXiv:1901.07488}}, \href {https://doi.org/10.1103/PhysRevD.99.096011} {\path{doi:10.1103/PhysRevD.99.096011}}.

\bibitem{He:2005sp}
L.~He, P.~Zhuang, {Phase structure of Nambu-Jona-Lasinio model at finite isospin density}, Phys. Lett. B 615 (2005) 93--101.
\newblock \href {http://arxiv.org/abs/hep-ph/0501024} {\path{arXiv:hep-ph/0501024}}, \href {https://doi.org/10.1016/j.physletb.2005.03.066} {\path{doi:10.1016/j.physletb.2005.03.066}}.

\bibitem{Andersen:2007qv}
J.~O. Andersen, L.~Kyllingstad, {Pion Condensation in a two-flavor NJL model: the role of charge neutrality}, J. Phys. G 37 (2009) 015003.
\newblock \href {http://arxiv.org/abs/hep-ph/0701033} {\path{arXiv:hep-ph/0701033}}, \href {https://doi.org/10.1088/0954-3899/37/1/015003} {\path{doi:10.1088/0954-3899/37/1/015003}}.

\bibitem{Xia:2013caa}
T.~Xia, L.~He, P.~Zhuang, {Three-flavor Nambu\textendash{}Jona-Lasinio model at finite isospin chemical potential}, Phys. Rev. D 88~(5) (2013) 056013.
\newblock \href {http://arxiv.org/abs/1307.4622} {\path{arXiv:1307.4622}}, \href {https://doi.org/10.1103/PhysRevD.88.056013} {\path{doi:10.1103/PhysRevD.88.056013}}.

\bibitem{Abuki:2008wm}
H.~Abuki, R.~Anglani, R.~Gatto, M.~Pellicoro, M.~Ruggieri, {The Fate of pion condensation in quark matter: From the chiral to the real world}, Phys. Rev. D 79 (2009) 034032.
\newblock \href {http://arxiv.org/abs/0809.2658} {\path{arXiv:0809.2658}}, \href {https://doi.org/10.1103/PhysRevD.79.034032} {\path{doi:10.1103/PhysRevD.79.034032}}.

\bibitem{Khunjua:2018jmn}
T.~G. Khunjua, K.~G. Klimenko, R.~N. Zhokhov, {Chiral imbalanced hot and dense quark matter: NJL analysis at the physical point and comparison with lattice QCD}, Eur. Phys. J. C 79~(2) (2019) 151.
\newblock \href {http://arxiv.org/abs/1812.00772} {\path{arXiv:1812.00772}}, \href {https://doi.org/10.1140/epjc/s10052-019-6654-2} {\path{doi:10.1140/epjc/s10052-019-6654-2}}.

\bibitem{Khunjua:2018sro}
T.~G. Khunjua, K.~G. Klimenko, R.~N. Zhokhov, {Dualities in dense quark matter with isospin, chiral, and chiral isospin imbalance in the framework of the large-N$_{c}$ limit of the NJL$_{4}$ model}, Phys. Rev. D 98~(5) (2018) 054030.
\newblock \href {http://arxiv.org/abs/1804.01014} {\path{arXiv:1804.01014}}, \href {https://doi.org/10.1103/PhysRevD.98.054030} {\path{doi:10.1103/PhysRevD.98.054030}}.

\bibitem{Khunjua:2017khh}
T.~G. Khunjua, K.~G. Klimenko, R.~N. Zhokhov, V.~C. Zhukovsky, {Inhomogeneous charged pion condensation in chiral asymmetric dense quark matter in the framework of NJL$_2$ model}, Phys. Rev. D 95~(10) (2017) 105010.
\newblock \href {http://arxiv.org/abs/1704.01477} {\path{arXiv:1704.01477}}, \href {https://doi.org/10.1103/PhysRevD.95.105010} {\path{doi:10.1103/PhysRevD.95.105010}}.

\bibitem{Ebert:2016hkd}
D.~Ebert, T.~G. Khunjua, K.~G. Klimenko, {Duality between chiral symmetry breaking and charged pion condensation at large $N_c$: Consideration of an NJL$_2$ model with baryon, isospin, and chiral isospin chemical potentials}, Phys. Rev. D 94~(11) (2016) 116016.
\newblock \href {http://arxiv.org/abs/1608.07688} {\path{arXiv:1608.07688}}, \href {https://doi.org/10.1103/PhysRevD.94.116016} {\path{doi:10.1103/PhysRevD.94.116016}}.

\bibitem{Ueda:2013sia}
H.~Ueda, T.~Z. Nakano, A.~Ohnishi, M.~Ruggieri, K.~Sumiyoshi, {QCD phase diagram at finite baryon and isospin chemical potentials in Polyakov loop extended quark meson model with vector interaction}, Phys. Rev. D 88~(7) (2013) 074006.
\newblock \href {http://arxiv.org/abs/1304.4331} {\path{arXiv:1304.4331}}, \href {https://doi.org/10.1103/PhysRevD.88.074006} {\path{doi:10.1103/PhysRevD.88.074006}}.

\bibitem{Mannarelli:2019hgn}
M.~Mannarelli, {Meson condensation}, Particles 2~(3) (2019) 411--443.
\newblock \href {http://arxiv.org/abs/1908.02042} {\path{arXiv:1908.02042}}, \href {https://doi.org/10.3390/particles2030025} {\path{doi:10.3390/particles2030025}}.

\bibitem{Andersen:2023ivj}
J.~O. Andersen, Q.~Yu, H.~Zhou, {Pion condensation in QCD at finite isospin density, the dilute Bose gas, and speedy Goldstone bosons}, Phys. Rev. D 109~(3) (2024) 034022.
\newblock \href {http://arxiv.org/abs/2306.14472} {\path{arXiv:2306.14472}}, \href {https://doi.org/10.1103/PhysRevD.109.034022} {\path{doi:10.1103/PhysRevD.109.034022}}.

\bibitem{Ayala:2023cnt}
A.~Ayala, A.~Bandyopadhyay, R.~L.~S. Farias, L.~A. Hern\'andez, J.~L. Hern\'andez, {QCD equation of state at finite isospin density from the linear sigma model with quarks: The cold case}, Phys. Rev. D 107~(7) (2023) 074027.
\newblock \href {http://arxiv.org/abs/2301.13633} {\path{arXiv:2301.13633}}, \href {https://doi.org/10.1103/PhysRevD.107.074027} {\path{doi:10.1103/PhysRevD.107.074027}}.

\bibitem{Lopes:2021tro}
B.~S. Lopes, S.~S. Avancini, A.~Bandyopadhyay, D.~C. Duarte, R.~L.~S. Farias, {Hot QCD at finite isospin density: Confronting the SU(3) Nambu\textendash{}Jona-Lasinio model with recent lattice data}, Phys. Rev. D 103~(7) (2021) 076023.
\newblock \href {http://arxiv.org/abs/2102.02844} {\path{arXiv:2102.02844}}, \href {https://doi.org/10.1103/PhysRevD.103.076023} {\path{doi:10.1103/PhysRevD.103.076023}}.

\bibitem{Adhikari:2019zaj}
P.~Adhikari, J.~O. Andersen, {QCD at finite isospin density: chiral perturbation theory confronts lattice data}, Phys. Lett. B 804 (2020) 135352.
\newblock \href {http://arxiv.org/abs/1909.01131} {\path{arXiv:1909.01131}}, \href {https://doi.org/10.1016/j.physletb.2020.135352} {\path{doi:10.1016/j.physletb.2020.135352}}.

\bibitem{Avancini:2019ego}
S.~S. Avancini, A.~Bandyopadhyay, D.~C. Duarte, R.~L.~S. Farias, {Cold QCD at finite isospin density: confronting effective models with recent lattice data}, Phys. Rev. D 100~(11) (2019) 116002.
\newblock \href {http://arxiv.org/abs/1907.09880} {\path{arXiv:1907.09880}}, \href {https://doi.org/10.1103/PhysRevD.100.116002} {\path{doi:10.1103/PhysRevD.100.116002}}.

\bibitem{Andersen:2015eoa}
J.~O. Andersen, N.~Haque, M.~G. Mustafa, M.~Strickland, {Three-loop hard-thermal-loop perturbation theory thermodynamics at finite temperature and finite baryonic and isospin chemical potential}, Phys. Rev. D 93~(5) (2016) 054045.
\newblock \href {http://arxiv.org/abs/1511.04660} {\path{arXiv:1511.04660}}, \href {https://doi.org/10.1103/PhysRevD.93.054045} {\path{doi:10.1103/PhysRevD.93.054045}}.

\bibitem{Mukherjee:2006hq}
S.~Mukherjee, M.~G. Mustafa, R.~Ray, {Thermodynamics of the PNJL model with nonzero baryon and isospin chemical potentials}, Phys. Rev. D 75 (2007) 094015.
\newblock \href {http://arxiv.org/abs/hep-ph/0609249} {\path{arXiv:hep-ph/0609249}}, \href {https://doi.org/10.1103/PhysRevD.75.094015} {\path{doi:10.1103/PhysRevD.75.094015}}.

\bibitem{Chen:2024cxh}
Y.~Chen, M.~Ding, D.~Li, K.~Bitaghsir~Fadafan, M.~Huang, {Pion Condensation and Pion Star from Holographic QCD} (8 2024).
\newblock \href {http://arxiv.org/abs/2408.17080} {\path{arXiv:2408.17080}}.

\bibitem{Brandt:2022fij}
B.~B. Brandt, F.~Cuteri, G.~Endr\"odi, {Equation of state and Taylor expansions at nonzero isospin chemical potential}, PoS LATTICE2022 (2023) 144.
\newblock \href {http://arxiv.org/abs/2212.01431} {\path{arXiv:2212.01431}}, \href {https://doi.org/10.22323/1.430.0144} {\path{doi:10.22323/1.430.0144}}.

\bibitem{Brandt:2022hwy}
B.~B. Brandt, F.~Cuteri, G.~Endrodi, {Equation of state and speed of sound of isospin-asymmetric QCD on the lattice}, JHEP 07 (2023) 055.
\newblock \href {http://arxiv.org/abs/2212.14016} {\path{arXiv:2212.14016}}, \href {https://doi.org/10.1007/JHEP07(2023)055} {\path{doi:10.1007/JHEP07(2023)055}}.

\bibitem{Stiele:2013pma}
R.~Stiele, E.~S. Fraga, J.~Schaffner-Bielich, {Thermodynamics of (2+1)-flavor strongly interacting matter at nonzero isospin}, Phys. Lett. B 729 (2014) 72--78.
\newblock \href {http://arxiv.org/abs/1307.2851} {\path{arXiv:1307.2851}}, \href {https://doi.org/10.1016/j.physletb.2013.12.053} {\path{doi:10.1016/j.physletb.2013.12.053}}.

\bibitem{Adhikari:2018cea}
P.~Adhikari, J.~O. Andersen, P.~Kneschke, {Pion condensation and phase diagram in the Polyakov-loop quark-meson model}, Phys. Rev. D 98~(7) (2018) 074016.
\newblock \href {http://arxiv.org/abs/1805.08599} {\path{arXiv:1805.08599}}, \href {https://doi.org/10.1103/PhysRevD.98.074016} {\path{doi:10.1103/PhysRevD.98.074016}}.

\bibitem{Ayala:2023mms}
A.~Ayala, B.~S. Lopes, R.~L.~S. Farias, L.~C. Parra, {Describing the speed of sound peak of isospin-asymmetric cold strongly interacting matter using effective models} (10 2023).
\newblock \href {http://arxiv.org/abs/2310.13130} {\path{arXiv:2310.13130}}.

\bibitem{Carlomagno:2024xmi}
J.~P. Carlomagno, D.~Gomez~Dumm, N.~N. Scoccola, {Cold isospin asymmetric baryonic rich matter in nonlocal NJL-like models}, Phys. Rev. D 109~(9) (2024) 094041.
\newblock \href {http://arxiv.org/abs/2402.13842} {\path{arXiv:2402.13842}}, \href {https://doi.org/10.1103/PhysRevD.109.094041} {\path{doi:10.1103/PhysRevD.109.094041}}.

\bibitem{Carlomagno:2021gcy}
J.~P. Carlomagno, D.~G. Dumm, N.~N. Scoccola, {Isospin asymmetric matter in a nonlocal chiral quark model}, Phys. Rev. D 104~(7) (2021) 074018.
\newblock \href {http://arxiv.org/abs/2107.02022} {\path{arXiv:2107.02022}}, \href {https://doi.org/10.1103/PhysRevD.104.074018} {\path{doi:10.1103/PhysRevD.104.074018}}.

\bibitem{Abbott:2023coj}
R.~Abbott, W.~Detmold, F.~Romero-L\'opez, Z.~Davoudi, M.~Illa, A.~Parre\~no, R.~J. Perry, P.~E. Shanahan, M.~L. Wagman, {Lattice quantum chromodynamics at large isospin density}, Phys. Rev. D 108~(11) (2023) 114506.
\newblock \href {http://arxiv.org/abs/2307.15014} {\path{arXiv:2307.15014}}, \href {https://doi.org/10.1103/PhysRevD.108.114506} {\path{doi:10.1103/PhysRevD.108.114506}}.

\bibitem{Abbott:2024vhj}
R.~Abbott, W.~Detmold, M.~Illa, A.~Parre\~no, R.~J. Perry, F.~Romero-L\'opez, P.~E. Shanahan, M.~L. Wagman, {QCD constraints on isospin-dense matter and the nuclear equation of state} (6 2024).
\newblock \href {http://arxiv.org/abs/2406.09273} {\path{arXiv:2406.09273}}.

\bibitem{Brandes:2023hma}
L.~Brandes, W.~Weise, N.~Kaiser, {Evidence against a strong first-order phase transition in neutron star cores: Impact of new data}, Phys. Rev. D 108~(9) (2023) 094014.
\newblock \href {http://arxiv.org/abs/2306.06218} {\path{arXiv:2306.06218}}, \href {https://doi.org/10.1103/PhysRevD.108.094014} {\path{doi:10.1103/PhysRevD.108.094014}}.

\bibitem{He:2005nk}
L.-y. He, M.~Jin, P.-f. Zhuang, {Pion superfluidity and meson properties at finite isospin density}, Phys. Rev. D 71 (2005) 116001.
\newblock \href {http://arxiv.org/abs/hep-ph/0503272} {\path{arXiv:hep-ph/0503272}}, \href {https://doi.org/10.1103/PhysRevD.71.116001} {\path{doi:10.1103/PhysRevD.71.116001}}.

\bibitem{Carignano:2016lxe}
S.~Carignano, L.~Lepori, A.~Mammarella, M.~Mannarelli, G.~Pagliaroli, {Scrutinizing the pion condensed phase}, Eur. Phys. J. A 53~(2) (2017) 35.
\newblock \href {http://arxiv.org/abs/1610.06097} {\path{arXiv:1610.06097}}, \href {https://doi.org/10.1140/epja/i2017-12221-x} {\path{doi:10.1140/epja/i2017-12221-x}}.

\bibitem{Chiba:2023ftg}
R.~Chiba, T.~Kojo, {Sound velocity peak and conformality in isospin QCD}, Phys. Rev. D 109~(7) (2024) 076006.
\newblock \href {http://arxiv.org/abs/2304.13920} {\path{arXiv:2304.13920}}, \href {https://doi.org/10.1103/PhysRevD.109.076006} {\path{doi:10.1103/PhysRevD.109.076006}}.

\bibitem{Kojo:2024sca}
T.~Kojo, D.~Suenaga, R.~Chiba, {Isospin QCD as a Laboratory for Dense QCD}, Universe 10~(7) (2024) 293.
\newblock \href {http://arxiv.org/abs/2406.11059} {\path{arXiv:2406.11059}}, \href {https://doi.org/10.3390/universe10070293} {\path{doi:10.3390/universe10070293}}.

\bibitem{Mu:2010zz}
C.-f. Mu, L.-y. He, Y.-x. Liu, {Evaluating the phase diagram at finite isospin and baryon chemical potentials in the Nambu-Jona-Lasinio model}, Phys. Rev. D 82 (2010) 056006.
\newblock \href {https://doi.org/10.1103/PhysRevD.82.056006} {\path{doi:10.1103/PhysRevD.82.056006}}.

\bibitem{Fujimoto:2023mvc}
Y.~Fujimoto, {Enhanced contribution of the pairing gap to the QCD equation of state at large isospin chemical potential}, Phys. Rev. D 109~(5) (2024) 054035.
\newblock \href {http://arxiv.org/abs/2312.11443} {\path{arXiv:2312.11443}}, \href {https://doi.org/10.1103/PhysRevD.109.054035} {\path{doi:10.1103/PhysRevD.109.054035}}.

\bibitem{Adhikari:2019mdk}
P.~Adhikari, J.~O. Andersen, P.~Kneschke, {Two-flavor chiral perturbation theory at nonzero isospin: Pion condensation at zero temperature}, Eur. Phys. J. C 79~(10) (2019) 874.
\newblock \href {http://arxiv.org/abs/1904.03887} {\path{arXiv:1904.03887}}, \href {https://doi.org/10.1140/epjc/s10052-019-7381-4} {\path{doi:10.1140/epjc/s10052-019-7381-4}}.

\end{thebibliography}

\end{document}